\documentclass[a4,12pt,fleqn]{article}
\usepackage{epsfig,times,latexsym,enumerate,lscape,float,flafter,amsmath,amssymb,graphicx}

\newcommand{\prd}{Phys. Rev. D }
\newcommand{\mnras}{MNRAS }
\newcommand{\cqg}{Class. Quantum Grav. }

\begin{document}

\title{On the independent points in the sky for the search of periodic gravitational wave}
\author{S.K. Sahay\thanks{email: ssahay@iucaa.ernet.in, ssahay@bits-goa.ac.in}\\
{\small Birla Institute of Technology and Science-Pilani, Goa Campus, Physics Group, Zuarinagar, Goa - 403726, India.}}

\date{}
\maketitle

\begin{abstract} We investigate the independent points in the sky require
to search the periodic gravitational wave, assuming the noise power spectral density to be flat. We have made an analysis with different initial azimuth of 
the Earth for a week data set. The analysis shows 
significant difference in the independent points in the sky for the search. We numerically obtain an approximate relation to make trade-off between computational cost and sensitivities. We also discuss the feasibility of the coherent search in small frequency band in reference to advanced LIGO. 
\end{abstract}

\section{Introduction}
The first generation of laser interferometric gravitational wave observatory (LIGO) [Abramovici (1992)] is in operation. 
However, its has been planned that by the end of the initial LIGO science run it would undergo upgrades to significantly improve the sensitivity. Now referred to as advanced LIGO (formerly LIGO II)  [Weinstein, 2002]. The detector will sweep their broad 
quadrupolar beam pattern across the sky as the earth moves. Hence the data 
analysis system will have to carry
out all sky searches for its sources. In this, search of the PGW without a priori knowledge appears to 
computationally quite demanding even by the standard computers
expected to available in the near future. It appears that due to limited computational resource it will be not feasible to perform all sky all frequency 
search   
in the months/year data set. However, if advanced LIGO achieve its design sensitivity $\sim 10^{-23}$ or better (Weinstein, 2002), then it may be feasible to perform all sky search for a day to week data set in small frequency
band for the sources emitting signal of amplitude $\gtrsim 10^{-26}$. The 
search of the potential sources may be more significant, if done in 
the frequency 
band where most of the Pulsars are detected by other means. Also, the choice of sophisticated, optimal data analysis methods and a clever programming is also
integral part to search the signal buried in the noise with the available computation power.  

\par The current status of the 
search indicates that its important to detect the gravitational waves (GW) rather finding the source 
location more accurately. Hence, one would like to do minimum Doppler 
correction or/and to make the search 
templates. In reference to the all sky search, Schutz (1991), has introduced the 
concept of patch in the 
sky as the region of space throughout which the required Doppler correction 
remains the same. He roughly estimated the number of patches required and shown that  
for $10^7$ sec. observation data set of one KHz signal would be about $1.3 
\times 10^{13}$. This also indicates the bank of templates require 
for the coherent all sky search. The size of the patch may be increased, hence reducing the number of points require for the Doppler correction,
by manipulating the output of the detector. Which in turn demand the detail investigation 
of the parameters affecting the phase of the modulated signal. In this, the initial azimuth of the Earth play a vital role in the modulation of the signal, particularly for the analysis of day to week data set. Hence, in the next section, we incorporate the 
initial azimuth of the Earth in the Fourier transform (FT) obtained by 
Srivastava and Sahay (2002a,b) for arbitrary observation time. 

\par The limited computation power make to search the signal in the short observation data set, may be a day/week, so in section 3,  employing the  concept of fitting factor (Apostolatos, 1995), we investigate the independent points in the sky ($N_{sky}$) require for an all sky 
search for a week data set with different initial azimuth of the Earth, 
assuming the noise power spectral density to be flat. Also, we numerically obtain an approximate relation to make trade-off between computational cost and sensitivities. In section 4, we discuss the feasibility of the coherent search in reference to advanced LIGO. Section 5 contains the conclusions of the paper.

\section{Fourier transform}

The FT analysis of the frequency modulated PGW signal has been done by 
Srivastava and Sahay (2002a,b) by taking account the effects arising due to 
the rotational as well 
as orbital motion of the Earth. However, they have neglected an important 
parameter, the initial azimuth of the Earth, which affects significantly in the spacing of the parameters space for an all sky search. To obtain the FT with taking account the Earth initial azimuth, we rewrite the phase of the received PGW signal of frequency $f_\circ$ at 
time $t$ given by them and may be written as

\begin{eqnarray}
 \Phi (t) & = & 2\pi f_\circ t + {\cal Z}\cos (a\xi_{rot} - \sigma ) + {\cal N}\cos (\xi_{rot} - \delta ) - {\cal M}
\end{eqnarray}
\newpage
\noindent where \\
\begin{equation}
\left.\begin{array}{lcl}
\vspace{0.2cm}
{\cal M}& = & \frac{2\pi f_\circ}{c}\left(R_{se}\sin\theta\cos\sigma \sqrt{{\cal P}^2 + {\cal Q}^2 }\cos\delta\right) , \\
\vspace{0.2cm}
{\cal Z}& = & \frac{2\pi f_\circ}{c} R_{se}\sin\theta\, ,\\
\vspace{0.2cm}
{\cal N}&= & \frac{2\pi f_\circ}{c} \sqrt{{\cal P}^2 + {\cal Q}^2 } \, ,\\
\vspace{0.2cm}
{\cal P}& = & R_e\sin\alpha (\sin\theta\sin\phi\cos\epsilon  + \cos\theta \sin\epsilon )\, ,\\
\vspace{0.2cm}
{\cal Q}& = & R_e \sin\alpha\sin\theta\cos\phi \, ,\\
\vspace{0.2cm}
\sigma &=& \phi - \beta_{orb}\, , \quad \delta = \tan^{- 1}\frac{{\cal P}}{{\cal Q}} - \beta_{rot}\, , \\
\vspace{0.2cm}
a &= & w_{orb}/w_{rot}\; \approx \; 1/365.26, \; w_{orb}t \;  = \; a\xi_{rot} , \\
\vspace{0.2cm}
{\bf\textstyle n} & = & \left(\sin\theta\cos\phi , \; \sin\theta\sin\phi\, , \;\cos\theta\right), \; \xi_{rot} \; =\;  w_{rot}t \\
\end{array} \right\}
\end{equation}

\noindent  where $\theta$, $\phi$, $R_{e}$, $R_{se}$, $w_{orb}$ $w_{rot}$, $\alpha$ and 
$\epsilon$ represent 
respectively the celestial co-latitude, longitude, Earth radius, average distance between Earth centre from 
the origin of SSB frame, orbital and rotational angular velocity of the 
Earth, co-latitude of the detector and obliquity of the ecliptic. Here $\beta_{orb}$ and $\beta_{rot}$ are the initial azimuth of 
the Earth and detector respectively.\\ 

\noindent  To estimate $N_{sky}$, it is sufficient to consider either of the two polarisation of the signal given as,
\begin{equation}
h_+(t) = h_{\circ_+}\cos [\Phi (t)] 
\label{eq:hpt}
\end{equation}
\begin{equation}
h_\times (t) = h_{\circ_\times}\sin [\Phi (t)]
\label{eq:hct}
\end{equation}
\noindent hence, we consider the `$+$' polarisation of amplitude unity, given as
\begin{equation}
h(t) = \cos[\Phi (t)]
\label{eq:cosphit}
\end{equation}

\noindent where $h_{\circ_+}$, $h_{\circ_\times}$ are constant amplitude of the two 
polarizations.\\

Now let us assume $h(t)$ to be given on finite time interval $[0,T_{obs}]$ assumed to be the observation period. Now its straight forward to obtain FT in the similar way as obtain by Srivastava and Sahay (2002b), and may be given as

\begin{eqnarray}
\tilde{h}(f)& =& \int_0^{T_{obs}} \cos[\Phi (t)]e^{-i2\pi ft}dt  \nonumber \\
&& \simeq  \frac{\nu}{2 w_{rot}} \sum_{k  =  - 
\infty}^{k = \infty} \sum_{m = - \infty}^{m =  \infty} e^{ i {\cal A}}{\cal 
B}[ \tilde{{\cal C}} - i\tilde{{\cal D}} ] \; ; \;
\label{eq:hf}
\end{eqnarray}  

\noindent where
\begin{equation}
\left.\begin{array}{lcl}
\vspace{0.2cm}
\nu & = & \frac{f_\circ - f}{f_{rot}} \\
\vspace{0.2cm}
{\cal A}&  = &{(k + m)\pi\over 2} - {\cal M}  \\
\vspace{0.2cm}
{\cal B} & = & {J_k({\cal Z}) J_m({\cal N})\over {\nu^2 - (a k + m)^2}}\\
\vspace{0.2cm}
\tilde{{\cal C}} &=& \sin \nu\xi_\circ \cos ( a k \xi_\circ + m\xi_\circ - k \sigma - m \delta )\\
\vspace{0.2cm}
&&  - { a k + m \over \nu}\{\cos\nu\xi_\circ \sin ( a k \xi_\circ + m\xi_\circ - k \sigma - m \delta )+ \sin ( k \sigma + m \delta )\}\\
\vspace{0.2cm}
\tilde{{\cal D}} & = & \cos \nu\xi_\circ \cos ( a k \xi_\circ + m\xi_\circ - k \sigma - m \delta )   \\
\vspace{0.2cm}
&& + {k a + m \over \nu}\sin \nu \xi_\circ \sin ( a k \xi_\circ + m\xi_\circ - k \sigma - m \delta ) - \cos ( k \sigma + m \delta )\\
\vspace{0.2cm}
\xi_o & = & w_{rot}T_{obs}
\end{array} \right\}
\end{equation}

\noindent where J stands for the Bessel function of first kind. Using the symmetrical property of Bessel function we reduce the computation time appreciably by rewriting $\tilde{h}(f)$ as  
\begin{eqnarray}
\tilde{h}(f)&\simeq & { \nu \over w_{rot}}\left[ {J_\circ({\cal Z}) J_\circ({\cal N}) \over 2\nu^2}\left[ \{ \sin{\cal M} - \sin
({\cal M} - \nu\xi_\circ )\}\; \right.+   i\{ \cos{\cal M} - \cos ({\cal M} - \nu\xi_\circ
 )\} \right]\;+ \nonumber \\
&& J_\circ ({\cal Z})\sum_{m = 1}^{m = \infty} {J_m({\cal N})\over 
\nu^2 - m^2} \left[ ( {\cal Y} {\cal U} -  {\cal X} {\cal V} ) - i ( 
{\cal X} {\cal U} + {\cal Y} {\cal V} ) \right]\; + \nonumber \\ 
&& \left.\sum_{k = 1 }^{k = \infty}\sum_{m = -
\infty}^{m = \infty} e^{ i {\cal A}}{\cal B}\left(
\tilde{{\cal C}} - i\tilde{{\cal D}} \right)\right]\; ;
\label{eq:fm_code}
\end{eqnarray} 

\begin{equation}
\left.\begin{array}{ccl}
{\cal X}& =& \sin ({\cal M}  - m \pi/2 )\\
{\cal Y}& =& \cos ({\cal M} - m \pi/2 )\\
{\cal U}& =& \sin \nu\xi_\circ \cos m ( \xi_\circ - \delta ) - {m\over \nu}\left\{\cos
\nu\xi_\circ \sin m ( \xi_\circ - \delta ) - \sin m\delta\right\}\\
{\cal V}& =& \cos \nu\xi_\circ \cos m ( \xi_\circ - \delta ) + {m\over \nu}\sin 
\nu\xi_\circ \sin m ( \xi_\circ - \delta ) - \cos m\delta\\
\end{array}\right\}
\end{equation}

The transform contains double infinite series of Bessel function. 
However, for analysis the order of Bessel function required to compute 
$\tilde{h}(f)$ in the infinite series are given as (Srivastava and Sahay, 2002b). 
\begin{equation}
k\approx 3133.22 \times 10^3 \sin\theta \left(\frac{f_\circ}{1 kHz}\right)\, ,
\end{equation}
\begin{equation}
m\approx 134 \left(\frac{f_\circ}{1 kHz}\right).
\end{equation}

The accuracy and range of validity for large order and/or argument has been discussed by Chishtie et.al (2005).

\section{Independent points in the sky} 
The study of the independent points for an all sky search
has been made by many research workers [Schutz
(1991), Brady et. al. (1998), Brady and Creighton (2000), Jaranowski and  Kr\'olak (1999, 2001), Astone et.al (2002)] for the coherent and/or incoherent search. The coherent search means 
cross correlating the data with the bank of search templates. While incoherent search implies adding of the power spectra by dividing the data into $N$ subsets, performing a full search for each subset, and adds up the power spectra of the resulting 
searches. In this case, there is loss in S/N ratio by a factor of $\sqrt{N}$ in relation to coherent search as power spectra are added incoherently.
However, irrespective of the method of search, the optimal spacing in the $(\theta,\phi)$ parameters for an all sky search is a problem of interest. 

\par For the coherent search, one have to make bank of search templates to detect the 
signal. The bank of search 
templates are discrete set of signals from
among the continuum of possible signals. Consequently all the 
signals will not get detected with equal efficiency. However, it is possible to 
choose judiciously the set of templates so that all the signals of a given 
amplitude are
detected with a given minimum detection loss. Fitting factor ($FF$) is one of the  standard 
measure for deciding what class of wave form is good enough and 
quantitatively describes the
 closeness of the true signals to the
template manifold in terms of the reduction of S/N arising due to the cross 
correlation of 
a signal outside the manifold with the best matching templates lying inside the 
manifold, given as

\begin{eqnarray} 
{FF} & = & \frac{\langle h(f)|
h_T(f;\theta_T , \phi_T)\rangle}{\sqrt{\langle h_T(f;\theta_T , \phi_T )|h_T(f;
\theta_T , \phi_T )\rangle\langle h(f)|h(f)\rangle}}
\label{eq:ff1}
\end{eqnarray}

\noindent where $h(f)$ and $h_T(f; \theta_T , \phi_T)$ represent the
FTs of the actual signal and the templates respectively. The inner product of two 
signal $h_1$ and $h_2$ is defined as
\begin{eqnarray}
\langle h_1|h_2\rangle & =& 2\int_0^\infty \frac{\tilde{h}_1^*(f)\tilde{h}_2(f)
+ \tilde{h}_1(f)\tilde{h}_2^*(f)}{S_n(f)}df \nonumber \\
 & = &
4\int_0^\infty \frac{\tilde{h}_1^*(f)\tilde{h}_2(f)}{S_n(f)}df
\label{eq:ip}
\end{eqnarray}

\noindent where $^*$ denotes complex conjugation and $S_n(f)$ is the 
spectral noise density of the detector.

\par To estimate optimal spacing in the parameters space one have to careful investigate the parameters contain in the 
phase of the modulated signal. Hence we check  
the  
 effect for different  $\beta_{orb}$ in $ \tilde{h}(f)$ for a week data set for the LIGO detector at Livingston [the position and orientation of the detector 
can be found in Allen (1995)] of unit amplitude signal for $f_\circ = 50$ Hz and ($\theta,\phi$) $=$($\pi /18, \pi /4$). We take the ranges of $k$ and $m$ as 1 to 27300 and -15 
to 15 respectively. We found that Earth azimuth affects the FM spectrum severely. Hence, for the coherent search, we investigate its effect for the number of independent of points in the sky.

\par
\begin{figure}[h]
\centering
\epsfig{file=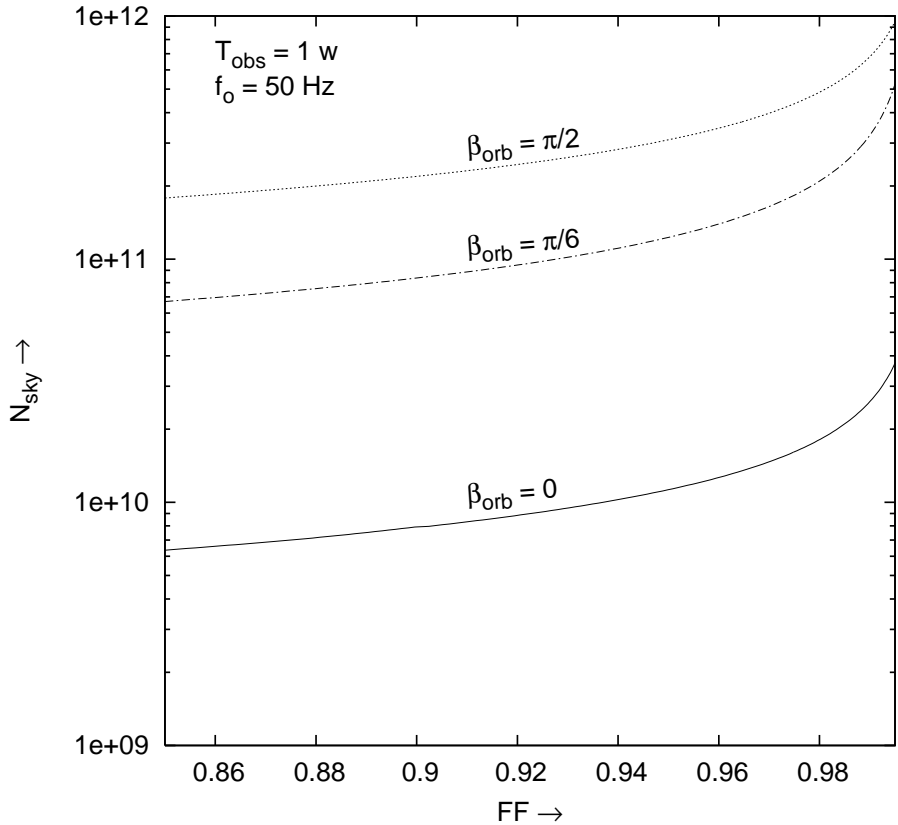,height=16.0cm,angle=-90}
\vspace{1.0cm}
\caption{$N_{sky}$ with FF at different $\beta_{orb}$.}
\label{fig:Nsky}
\end{figure}

\begin{figure}[h]
\centering
\epsfig{file=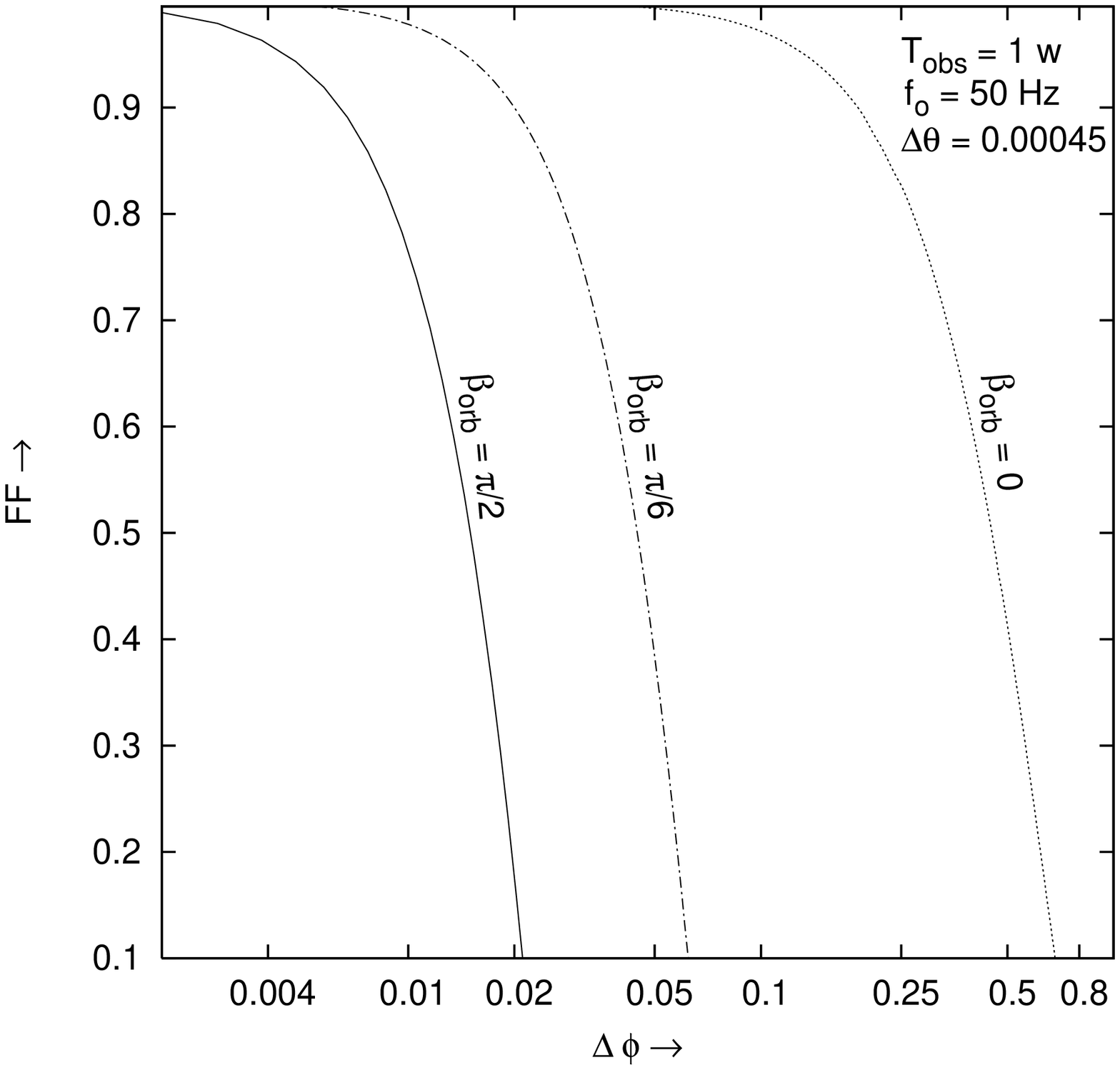,height=16.0cm,angle=-90}
\caption{FF with $\bigtriangleup\phi$ at different $\beta_{orb}$.}
\label{fig:deltaphi}
\end{figure}

 To estimate $N_{sky}$, we consider the LIGO detector at Livingston, receive a PGW signal of 
frequency $f_\circ =50$ Hz from a source 
located at $(\theta , \phi) = (1^\circ,45^\circ)$. First, we chosen the data set such that $\beta_{orb} =0$ at $t=0$. In this case we take
the ranges of $k$ and 
$m$ as $1$ to $2800$ and $-15$ to $15$ respectively and bandwidth equal to  
$50 \pm 3.28 \times 10^{-4}$ Hz for the integration. Now, we select 
the spacing $\bigtriangleup\theta = 0.45 \times 10^{-4}$, thereafter we maximize over $\phi$ by introducing 
spacing $\bigtriangleup\phi$ in the so obtained $N_{sky}$ and determine 
the resulting $FF$. In similar manner we obtain $N_{sky}$ for $\beta_{orb}=\pi/6$ and $\pi/2$. The results obtained are shown in the Fig.~(\ref{fig:Nsky}). We also plot the templates spacing $\bigtriangleup\phi$ in the $\phi-$parameter with FF shown in the Fig.~(\ref{fig:deltaphi}). Interestingly, the nature
of these curves are similar. we have obtained a best fit of the graphs, given as
\begin{equation}
N_{sky} = 10^{18} [a_\circ +a_1x -a_2x^2 +a_3x^3 -a_4x^4 +a_5x^5 -a_6x^6 +a_7x^7]; \quad 0.85 \le x \le 0.995;\end{equation}
\begin{equation}
FF = b_\circ + b_1y - b_2y^2 + b_3y^3;\quad 0.037 \le y \le 0.69;
\label{eq:ff}
\end{equation}
\noindent where $a_\circ..a_7$ and $b_\circ..b_3$ are constants as given in
Table 1 and 2 respectively.\\

\begin{table}[h]
\label{coefficients1}
\centering
{\begin{tabular}{ccccccccc}\\
\hline
$\beta_{orb}$ & $a_\circ$ & $a_1$ & $a_2$ & $a_3$ & $a_4$ & $a_5$ & $a_6$ & $a_7$ \\
\hline
&&&&&&\\
$0$&$-1.71829$& $13.1670 $ & $43.2234$ & $78.7956 $&$86.1505$&$56.4921$&$20.5716 $&$3.20919$\\
&&&&&&\\
$\pi/6$&$-35.7699$& $273.770$ & $897.636$ & $1634.42$&$1784.84 $&$1168.99$&$425.178$&$66.2488$\\
&&&&&&\\ 
$\pi/2$ &$-40.7053$& $311.849$ & $1023.49$ & $1865.42 $&$2039.12 $&$1336.86$&$486.720$&$75.9142$\\ 
&&&&&&\\\hline
\end{tabular}}
\caption{Coefficients of the best fit graphs obtained for the $N_{sky}$ with $FF$.}
\end{table}

\begin{table}[hbt]
\label{coefficients2}
\centering
{\begin{tabular}{ccccc}\\
\hline
$\beta_{orb}$ & $b_\circ$ & $b_1$ & $b_2$ & $b_3$\\
\hline
 0 & $0.994123$ & $0.149451$ & $3.94494$ & $2.64819$\\
&&&&\\
$\pi/6$ & $0.993544$ & $2.00186$ & $366.375$ &$1651.88$\\
&&&&\\
$\pi/2$ &$0.996774$ & $2.11099$ &$2914.21$ &$37817.3$\\
\hline
\end{tabular}}
\caption{Coefficients of the best fit graphs obtained for the $FF$ with $\bigtriangleup\phi$.}
\end{table}

In view of the above investigation, the 
spacing $\bigtriangleup\phi $ in the $\phi-$parameter may be
expressed as   
\begin{equation}
\bigtriangleup\phi = {\cal G}(FF, f_\circ, \theta ,\phi ,T_{obs}, \beta_{orb})
\label{eq:phifn}
\end{equation}

Equation~(\ref{eq:ff}) is a third order polynomial, hence complicates the solution. However, one would like to do data analysis for $FF > 0.90$. Therefore from the figure~(\ref{fig:deltaphi})  we obtain a very good dependence of $FF$ for minimum $N_{sky}$ and may be given as 
\begin{equation}
FF = - 2.85657 \bigtriangleup\phi^2 + 0.00230059 \bigtriangleup\phi  + 1.00018 
\label{eq:phiff}
\end{equation}

\noindent From Eqs.~(\ref{eq:phifn}), and~(\ref{eq:phiff}), we may write
\begin{eqnarray}
&{\cal F}(FF,50,1^\circ,45^\circ,1w,\beta_{orb}) \approx  4.02684 \times 10^{-4} \pm 0.591667 \sqrt{1.00018 - FF} 
\label{eq:tradeoff}
\end{eqnarray}

Above equation can be relevant to make trade-off between computational costs and sensitivities i.e. for the selected $FF$ one can estimate $N_{sky}$. However,  there is no unique choice for it. 
Here we are interested in the estimation of $N_{sky}$ such that the 
spacing is maximum resulting into the least number of points require for the search. As mentioned earlier,  
there is stringent requirement on reducing computational costs. Accordingly, 
there is serious
need of adopting some procedure/formalism to achieve this. For example, one
may adopt the method of
hierarchical search. 

\section{Computational costs}
In view of the above analysis it will be interesting to know the feasibility 
of the all sky search with the target sensitivity of the advanced LIGO. The computational costs of the data analysis basically depends on the floating point operations (flops) require to perform the Fast Fourier transform (FFT). Hence in terms of FFT, the flops for the data reduction upto frequency $f$ for $T_{obs}$ of the interferometer output may be given as (Press et al., 1986)
\begin{equation}
N_{flops}= 2fT_{obs}log_2(2f_{max}T_{obs})
\end{equation}

\noindent Now for the given mismatch ($FF$), if $N_p$ is the number of 
independent points to perform all sky search, then the flops will be
\begin{equation}
N_{flops}(FF,N_p)= 2fT_{obs}N_plog_2(2f_{max}T_{obs})
\end{equation}

Hence, for the mismatch of 3\% and without manipulating the data in reference to
$\beta_{orb}$, the flops for the search of PGW signal upto 50 HZ in the week data set 
will be  $2.38 \times 10^{19}, 2.58 \times 10^{20}$ and $6.23 \times 10^{20}$ 
for $\beta_{orb} = 0,\pi/6$ and $\pi/2$ respectively, assuming other operation need negligible flops compare to FFT. However, the lower cut off frequency of the LIGO I/II is 10/40 Hz. Hence the analysis shall be done above the lower cut off. Also, the search will be more significant if one perform in the most sensitive band of the detector. In this, if one would like to perform all sky search in a small band say 5 Hz then the minimum flops for the on-line analysis 
(a week data gets analyzed in a $\sim$ week time) will be
 $3.44 \times 10^{12}$. The flops require may be further reduce, if one perform hierarchal search. Hence, it may be feasible to perform limited frequency all sky search of signal amplitude $\gtrsim 10^{-26}$ in the output of such a sensitive detector with a $\sim$Tflops computer. 

\section{Conclusions}

In this paper we have incorporated the initial azimuth of the Earth in the 
FT of the frequency modulated PGW signal and investigated its effect in the independent points in the sky require for the search of PGW. We found that the $N_{sky}$ for the search in the output of one week data set  varies significantly with $\beta_{orb}$. For the case investigated here, we observe that  for $FF=0.97$ approximately $1.53 \times 10^{10}$, $1.66 \times 10^{11}$ and $4.0 \times 10^{11}$ 
$N_{sky}$ will be require when  
 $\beta_{orb} = 0, \pi/6$ and $\pi/2$ respectively. 
Hence, the analysis may be useful to reduce the computational cost for a coherent all sky search. However, also the inspection of the phase of the modulated signal reveals that reduction in $N_{sky}$ depends on time scale of integration, 
shorter the $T_{obs}$ more the difference in $N_{sky}$. 

\par The reduction in $N_{sky}$ is large, so we studied the feasibility of all sky search in reference to advanced LIGO and found that in the band of 5 Hz one may perform on-line all sky search of the PGW signal of amplitude $\gtrsim 10^{-26}$ with a $\sim$ Tflops computer. The relation given by equation~(\ref{eq:tradeoff}) may be useful to make trade-off between computational costs and sensitivities for the search of periodic gravitational waves. The issue to reduce the flops for the all sky search is a problem of interests and hence need more studies/investigations.\\

\section*{Acknowledgements}
I am thankful to Prof. D.C. Srivastava, Department of Physics, DDU Gorakhpur University, Gorakhpur for useful discussions. I thankfully acknowledge the computation facilities provided by IUCAA.

\end{document}